# Enhancing Performance of Cloud-based Software Applications with GraalVM and Quarkus

M. Šipek, D. Muharemagić, B. Mihaljević, and A. Radovan
Rochester Institute of Technology Croatia, Zagreb, Croatia
matija.sipek@mail.rit.edu, dino.muharemagic@mail.rit.edu,
branko.mihaljevic@croatia.rit.edu, aleksander.radovan@croatia.rit.edu

*Abstract* - Increased complexity of network-based software solutions and the ever-rising number of concurrent users forced a shift of the IT industry to cloud computing. Conventional network software systems commonly based on monolithic application stack running on costly physical single-purpose servers are affected by significant problems of resource management, computing power distribution, and scalability. Such implementation is restricting applications to be reduced to smaller, independent services that can be more easily deployed, managed, and scaled dynamically; therefore, embellishing environmental uniformity across development, testing, and production. Current cloud-based infrastructure frequently runs on containers placed in Kubernetes or Docker-based cluster, and the system configuration is considerably different compared to the environment prevailed with common virtualizations. This paper discusses the usage of GraalVM, a polyglot high-performance virtual machine for JVM-based and other languages, combined with new Kubernetes native Java tailored stacked framework named Quarkus, formed from enhanced Java libraries. Moreover, our research explores GraalVM's creation of native images using Ahead-Of-Time (AOT) compilation and Quarkus' deployment to Kubernetes. Furthermore, we examined the architectures of given systems, various performance variables, and differing memory usage cases within our academic testing environment and presented the comparison results of selected performance measures with other traditional and contemporary solutions.

## I. INTRODUCTION

An ever-increasing number of requirements and users demands ubiquitous availability of network-based software solutions, both server-based and serverless. Server applications need to be stable, responsive, and capable of elastically adjusting performance in order to provide well-timed and adequate complex services. Traditional service-oriented architecture (SOA) focuses excessively on using heavyweight middleware technologies, protocols, and formats such as Simple Object Access Protocol (SOAP) and Web Services Description Language (WSDL), thus becoming too interdependent and not loosely-coupled. These integrated solutions are not as flexible and problems within one part of the service can make the whole inaccessible and consequently inappropriate [1]. The current architectural shift to microservices emphasized scalability, autonomy, faster delivery, and lightweightness of individual independent services [2]. In order to fully use the advantages of microservices, they should be deployed within a cloud-based environment, using lightweight container technologies such as Docker[1] or Kubernetes[2].

One of the clouds' purposes is to set aside complex implementations presented by ununiformed IT-related impediments and help software developers focus on the implementation of core business functionalities while reducing the cost of deployment in comparison to traditional server-based applications. Cloud computing is a software deployment model that embodies several qualifications required by modern software, which allow omnipresence, rapid scalability, and on-demand access to network-based services [3]. As customer demand varies, consequently, various virtual and physical resources are reallocated in real time. In addition, one of the essential features leveraged in cloud systems is the ability of the service to track the number of consumers, bandwidth, server time, and memory processing, hence using a charge-per-use basis model of payment.

The most common cloud computing models are used in three forms: Software as a Service (SaaS), Platform as a Service (PaaS), and Infrastructure as a Service (IaaS) [4]; and the main deployment models: Public cloud, Private cloud, and Hybrid cloud [5]. Different software models are allowed to be deployed by various cloud providers such as leading IaaS vendors [3] Amazon Web Services (AWS), Microsoft Azure, Alibaba Cloud, Google Cloud Platform (GCP), Oracle Cloud, and IBM Cloud. Process Virtual Machines are virtualized and managed by a hypervisor on the hardware level [6]. The host processor is communicating with a thin layer of software systems also called Virtual Machine Management (VMM). As presented in Fig. 1., VMM manages sharing the physical resources between the several guest OSs running. While hardware virtualization emphasizes efficient resource utilization and lower cost, it still has limits in scalability, thus when a CPU overhead occurs on a particular VM, additional resources cannot be easily allocated to maintain the efficiency of a given server.

---

[1] Docker, https://www.docker.com/
[2] Kubernetes, https://kubernetes.io/
[3] Gartner 2019 Magic Quadrant for Cloud Infrastructure as a Service, Worldwide, Gartner, 2019

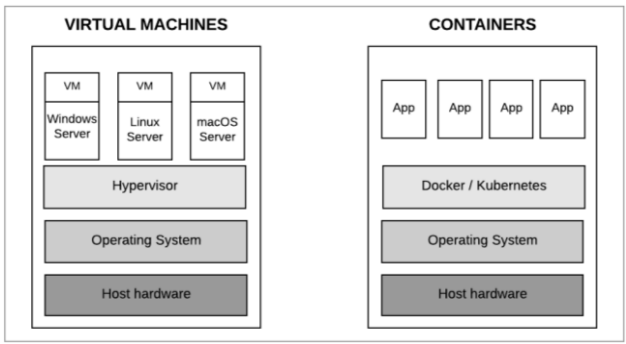

Figure 1. Common structure of containers and virtual machines

The separation of services and containerization is realized by operating system-level virtualization. The kernel allows multiple isolated instances to run simultaneously, while synchronously scaling computing resources. From the application's point of view, they are running on individual computers allowing common expensive resources to be shared.

As client requests and thus, overhead increases, the underlying cloud service providers spawn new work processes and increase computing resources in order to handle to incoming workload [7]. Initial warmup phase of VMs and the spawn of new workers in order to handle the request is named cold start, and it requires initialization of system service configurations and language runtime subsystems before handling requests. Cold start is usually a slow and unpredictable process [8], and this is even more enhanced in cloud environments. High-level object-oriented programming (OOP) languages running on Common Language Runtime (CLR) or Java VM are sometimes slower as they perform several expensive optimization and initialization processes such as class loading, profiling, and dynamic class loading [8].

In our previous research [9] we explored the essential aspects of the GraalVM[4], a high-performance universal polyglot virtual machine, which presented features with high industry potential such as embeddability and the usage of native images in serverless services. As defined in [10] and [11], usage of open source cloud in enterprise systems can excel in the utilization of several crucial business resources, and with the usage of GraalVM this facilitates further. Furthermore, in this research paper, we are exploring GraalVM's ahead-of-time (AOT) compilation mechanism of the Java code to a standalone native executable. In combination with Kubernetes and Java tailored stacked framework Quarkus [5] we tested and analyzed several performance variables, as well as elaborated common system components.

## II. GraalVM-based System Overview

### A. GraalVM system

GraalVM was developed as a part of OpenJDK's Graal Project [6] initiated to build a next-generation high-performance polyglot virtual machine that offers extensive ecosystem incorporating several important subsystems, including Graal Compiler, GraalVM Native Image mechanism, Truffle Language Implementation Framework and Sulong (LLVM), as presented in Figure 2. and in detail in [9].

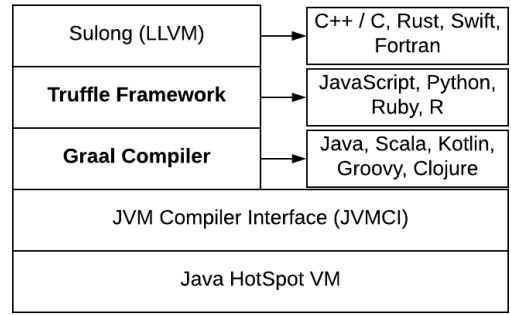

Figure 2. Simplified representation of GraalVM Architecture

GraalVM architecture starts with Java HotSpot VM at the bottom of the system stack. HotSpot VM's components that are not directly connected to the compilation process are being reused, comprising of Java Native Interface (JNI), interpreter, class libraries, class loading, as well as Garbage Collectors (GC) which efficiency is suitable and sufficient for the next generation of high-performance compilers [12], as tested in [13] and [14].

JVM Compiler Interface [7] (JVMCI) is connecting the Graal compiler with the underlying HotSpot VM in order to allow this new compiler written in Java to be used as the just-in-time (JIT) compiler, allowing JVM-based languages such as Java, Scala, Kotlin, Groovy and Clojure to be run, as well as a static, AOT compiler, used for creation of native images. Graal's modular architecture allows the reusability of common components; furthermore, taking the advantage of straightforwardness and affability of developing a compiler entirely in Java.

Truffle [8] is a framework used for implementing managed languages that are to be partially evaluated by the Graal compiler, thus allowing natural polyglotism within the system. It uses Abstract Syntax Tree (AST) as a simple way to interpret the program, but it is sometimes slower than expected and expensive. GraalVM recognizes interpreters written using the Truffle and is able to translate ASTs in highly optimized native executables. In order to eliminate overhead created by using cross-language scripts, any number of required ASTs can be joined together and interpreted as one [15].

Finally, on the top of the stack is Sulong, a compiler infrastructure engine for bitcode execution used for low-level programming languages interpretation on GraalVM, supporting common languages like C, C++, Fortran, Rust, and Swift, hence further extending GraalVM's versatility and interoperability.

---

## B. Native Image mechanism

In order to handle microservice requests within the cloud, the deployment of work processes requires an initialization subsystem, which would reduce the traditional system's startup time and memory footprint. One approach is Checkpoint/Restore In Userspace (CRIU) [9] used for freezing an application, saving it to persistant storage and when required, restoring and running application from that point. One of the issues with the checkpoint/restore concept is that it does not reduce the memory footprint that is systemic due to dynamic class loading and optimization, necessary for HotSpot VM memory needs for bytecode, dynamically compiled code, and metadata. GraalVM offers its Native Image mechanism completely written in Java, nevertheless offering applicability to other managed languages. A sequence of complementing mechanisms, such as point-to-analysis, build time initialization, heap snapshotting, and AOT compilation, achieves the preservation of integral Java benefits [7]. The complete system used for native image creation is named *image build time* and presented in Fig. 3.

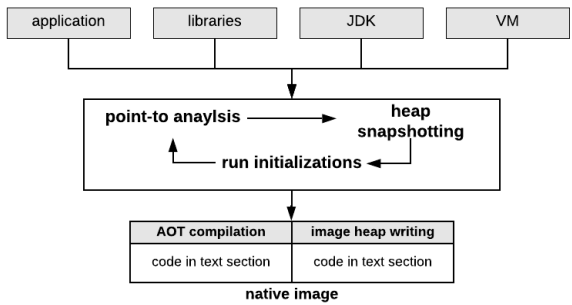

Figure 3. Image build time process

The aggressive static *point-to analysis* iteratively finds all the parts of the application which are reachable during runtime before a fixed point is reached. As analysis performs detailed processing, every field, method, and class is tracked individually, hence all parts of libraries on which the application depends are not included, but rather the necessary parts. The point-to analysis utilizes the frontend of the Graal compiler, using Graphbuilder [12] to turn Java bytecode into Graal Intermediate Representation (Graal IR) [7].

Execution of the code initialization starts when there are no more types to be added to type states, that is when the local fixed point is reached. The defining of the time at which classes are initialized in regards to image build time is up to the developer. Accordingly, information from methods, types or fields marked as reachable by point-to analysis facilitates the build of data structures that are optimized for reachable pieces of the application.

Afterward, the *heap snapshotting* mechanism searches for the objects created during the initialization and builds an object graph. Iterative analysis re-takes point-to analysis including heap snapshotting, until a point where the system does not find changes compared to the previous iteration. The latest graph instance is used to assume the role of image heap, and these objects are consequently serialized and placed into the data section of native image [7].

Lastly, the compiler used in the AOT Compilation mechanism is not the equivalent to the Graal compiler described in [9], yet it is its adjusted counter-part. The Graal compiler has modular architecture [16] and speculative compiler optimizations, which require additional consecutive deoptimizations to be run. However, deoptimizations cannot be used for the AOT compilation; henceforth, the compiler adapts point-to analysis results to optimize code performance, as presented in Fig. 3. Consequently, the part of the image heap was stacked during this process resulting in an executable image with a pre-populated heap.

## C. Quarkus framework

Native and serverless environments require additional adaption of containers and native images as larger applications consume more memory thus inducing significantly higher cost. Further enhancement of the Java system is supported by Quarkus, a Kubernetes Native Java stack framework, designed to work both with HotSpot VM and GraalVM. As Quarkus presents cloud-optimized Java libraries, constraints presented with the translation of Java into native images are reduced, thereby benefiting the image creation mechanism [17].

The configuration and automatization properties of Quarkus improves developers' workflow and productivity as Quarkus and its extensions can be easily configured. While creating an application, Quarkus' development mode offers automatic detection of changes and updates to the application. Thus, Quarkus opens HTTP request, reboots the application, presenting immediate changes accordingly allowing the developer to save time compared to common server solutions [17]. As one of the extensions, Quarkus uses Undertow[10], a flexible performant web server written in Java that further enhances the development process. It contributes to the usage of microservices as well as asynchronous programming in order to endure the modern standards of business applications which require high responsiveness.

The security of Quarkus applications is presented with extendability and the ability to reuse certain good practices. With the Quarkus core system, a security base extension is provided, which gives concrete implementation choices. A developer can choose to manually handle data by using properties file Security Extension. The next level of protection is the Java Database Connectivity (JDBC) Security Extension, setting data into developer maintained database. Nevertheless, the suggested practice is to delegate the security to a third-party provider that would handle key business components such as authentication, authorization, and user management.

Ultimately, Quarkus is a stable platform as it uses features provided by very stable ecosystems comprising of Hibernate ORM, Eclipse Vert.x, Netty, and RestEasy. It provides a unified configuration with a fast live reload, currently making it one of the leading platforms for effective Java solutions running in serverless environments.

---

[9] Checkpoint/Restart in Userspace (CRIU), https://www.criu.org/

[10] Undertow, http://undertow.io/

*D. Kubernetes platform*

Kubernetes is one of the popular containerization platforms used for microservice application bundling, enabling continuous development, deployment, and integration, as well as DevOps-like separation of concerns. This stable continuously building container image platform is able to create application container images at build time; however, as a result of running on a container level rather than on a traditional, hardware level with PaaS systems. Nevertheless, it provides key features for managing the deployment of a given system such as service discovery, storage orchestration, and load balancing with additional security and configuration mechanisms.

*E. Other competitors: Micronaut and Oracle Helidon*

Helidon[11] is an open-source set of Java libraries for writing microservices that offers two different programming models: Helidon SE and Helidon MP. Namely, Helidon SE is a microframework that supports the reactive programming model, while Helidon MP is an Eclipse MicroProfile based, allowing the Jakarta EE (former Java EE) style of programming and running microservices with CDI, JAX-RS, and JSON-P in a portable way.

Micronaut[12] is, similarly, a complete polyglot full-stack application framework for any type of application, while being focused on microservices and serverless applications. It uses AOT for all dependency and configuration injections for precomputing annotation metadata during compilation time, consequently removing all the metadata work at runtime. Hence, anything that is framework infrastructure is performed at compilation time to avoid higher memory and startup time at runtime.

III. THE BENCHMARK AND PRELIMINARY RESULTS

*A. The benchmark suite*

For this research we selected the Renaissance benchmark suite [13], composed of modern, real-world, concurrent, and object-oriented workloads that exercise various concurrency primitives of the JVM. This new open-source suite proposes a representative set of 21 benchmarks that bring into focus concurrency and parallelism paradigms, as well as newer language and JVM features, such as Java Lambdas, invokedynamic, and method handles [18], giving it the advantage over other benchmark suites like DaCapo[14] or SPECjvm2008[15]. Its main intention is to optimize JIT compilers, interpreters, and garbage collectors for tools such as profilers, debuggers, and static analyzers, by aggregating prevalent modern JVM workloads, including Big Data, Machine Learning, and functional programming. We selected a set of several benchmark tests from Renaissance Suite version 0.10.0 to gain a better point of view on the benefits of GraalVM and standalone native executables.

The benchmark tests are described in [18] and we selected the following in our testing process:

- *future-genetic* – genetic algorithm function optimization using the Jenetics library and futures
- *philosophers* – a variant of the dining philosophers problem using ScalaSTM
- *movie-lens* – movie recommender using the ALS algorithm
- *par-mnemonics* – solves the phone mnemonics problem using parallel JDK streams

This set of benchmark tests we found to be the most suitable for our academic testing environment while at the same time offering a wide spectrum of paradigms, enabling us to obtain relevant information in numerous scenarios.

Most commonly, benchmark tests have a warm-up stage, which takes either a certain number of iterations (Renaissance, DaCapo) or execution time (SPECjvm2008) in order for the JVM to stabilize. Only after that criterion is met we can obtain the credible benchmark results from the execution, called the steady-state.

We have also implemented a RESTful API system[16] featuring CRUD operations in several modern frameworks, including Spring Boot (release 2.2.4), Quarkus (version 1.1.1), and Micronaut (version 1.2.10). Each of the implementations operate with a MySQL database through the Hibernate framework on a table. The table data was randomly generated and consists of approximately 250 rows. The applications have a set of 3 endpoints exposed, namely a GET mapping to /products, which retrieves a collection of products, a GET mapping to /products/{id}, which retrieves a specific product determined by its unique identifier, and a POST mapping to /products which creates a new product. The REST applications have been hit with 100 requests before collecting the results and calculating the average result of both the Resident Set Size (RSS) and response times.

*B. Testing environment configuration*

The benchmark tests were obtained from our testing environment consisting of several different JDKs:

- OpenJDK 64-Bit Server VM (AdoptOpenJDK) (build 25.242-b08) in mixed mode OpenJDK version 1.8.0_242
- OpenJDK 64-Bit Server VM (AdoptOpenJDK) (build 13.0.1+9) in mixed mode with the same javac version
- OpenJDK 64-Bit Server VM GraalVM CE 19.3.1 (build 11.0.6+9-jvmci-19.3-b07) in mixed mode with OpenJDK version 11.0.6

The operating system utilized for testing was Mac OS X 10.15.2 x86_64, with 2.3 GHz Quad-Core Intel Core i5 and 8 GB 2133 MHz LPDDR3 RAM.

---

[11] Helidon, https://helidon.io/
[12] Micronaut, https://micronaut.io/
[13] Renaissance Suite, https://renaissance.dev/
[14] DaCapo benchmark suite, http://dacapobench.org/
[15] SPECjvm2008 benchmark suite, https://www.spec.org/jvm2008/
[16] RESTful API source code, https://github.com/dmuharemagic/rest-api-benchmark-mipro

## C. Selected test cases and discussion of results

Our preliminary test results, as presented in Fig. 4., were gathered in the steady-state after the warm-up has occurred and the Renaissance benchmark tests were stabilized, thus recording and observing the best execution times. The figure presents the average test time expressed in milliseconds (ms), and each benchmark test was performed in 250 iterations. The tests were run on OpenJDK 8, OpenJDK 13, and Graal CE 19.3.1 environments, with their individual respective versions being referenced in the previous segment.

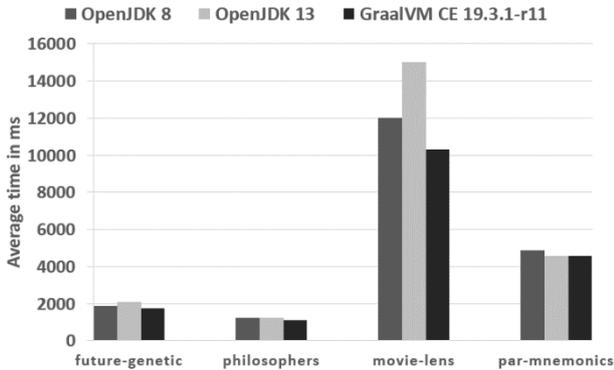

Figure 4. Preliminary benchmark test results for selected Renaissance benchmarks on Graal CE, OpenJDK 8, and OpenJDK 13

In the scenario of running the *movie-lens* benchmark test, Graal CE outperforms OpenJDK 13 by as much as 31.22% in terms of the decreased amount of running time. When speaking of *future-genetic* and *par-mnemonics* benchmarks, Graal CE performs better by a small difference in comparison to both OpenJDKs 8 and 13. Regarding the *future-genetic* benchmark test, Graal CE seems to surpass OpenJDK 8 by 6.95% and OpenJDK 13 by 15.18% in terms of speed.

Preliminary test results illustrated in Table 1. demonstrate the overall advantage of running native images in comparison with a traditional cloud-native stack. When measuring the memory footprint of our RESTful applications, in order to obtain complete results, we measure RSS, rather than the JVM heap size itself. Besides allocating native machine memory for the Heap, the total allocated memory for an application may also include class metadata, thread stacks, compiled code and garbage collection (varying on the JVM implementation). A standalone executable produced by GraalVM shows an 82.4% decrease in the usage of RSS memory in comparison to the same application implemented in Spring Boot running OpenJDK 13. Likewise, it is interesting to see that Micronaut outperforms Quarkus when running OpenJDK 13 by 22.92% in terms of memory allocation.

TABLE I. RESIDENT SET SIZE (RSS) IN SIDE-BY-SIDE COMPARISON LISTING COMMON MODERN FRAMEWORKS

| Benchmark | Resident Set Size (RSS) in kilobytes |
|---|---|
| Quarkus + Native (via GraalVM) | 31404 |
| Quarkus + OpenJDK 13 | 171940 |
| Micronaut + OpenJDK 13 | 132536 |
| Spring Boot + OpenJDK 13 | 178484 |

Similarly, the results presented in Figure 5. manifest the massive gain in response time when querying our RESTful API for a collection of objects. Quarkus running as a standalone native image produced by GraalVM runs 81.74% faster than the Spring Boot application with the same implementation. Quarkus implementation running on OpenJDK 13 runs 17.65% faster than the same Spring Boot application. Once again, Micronaut seems to outperform by 16.82% in terms of faster response times when running OpenJDK 13.

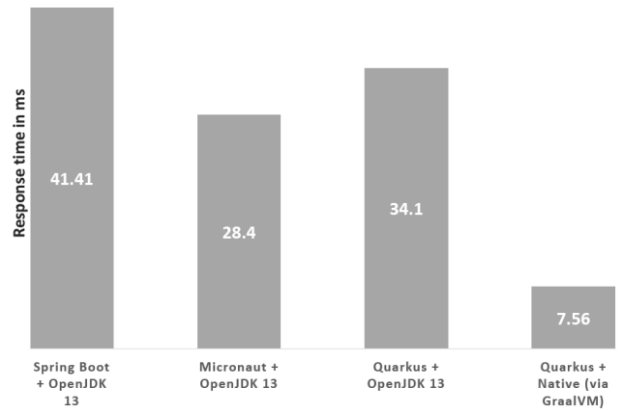

Figure 5. Response time in milliseconds obtained from a custom benchmark test (featuring a small RESTful API with basic CRUD operations) in selected frameworks

## IV. FUTURE WORK

This paper presents our continuous research [9] related to GraalVM, and in this paper we focused more on an AOT compiler. We believe that more features will further improve its extensibility and thus, industry usage. Serverless and microservice-based software systems are becoming the standard and will possibly take over the majority of network-based applications. Eventually, more focus will be needed in security aspects, as well as hardening the newly created system for improved resilience. As noted above, there are several competitors which we intend to compare in more detail together with OpenJDK 14 (in Early Access) and others. Additionally, other AOT compilers and cloud providers with adding new features, such as JVM-level threads named fibers, try to further speed up our system, and we plan to cover them in our future research.

## V. CONCLUSION

GraalVM's progress though years is visible and noticeable. In our first research [9] we mainly focused on testing and comparing GraalVM to the standard JDK version without deployment to server environments results confirmed our assumption and GraalVM's performance was surpassing competitor's. In this paper we presented preliminary results of our tests conducted considering GraalVM's native image compilation performance in combination with Quarkus gave good results, showing significantly reduced RSS.

GraalVM preliminary results on newly created benchmarking tool Renaissance, which focuses on increased service workload and asynchronous programming practices, slightly exceeds the current competitors. Since version 19 was released, we consider GraalVM ready for usage in the production environments, additionally as Oracle labs claim its stability and maturity, particularly in GraalVM EE.